\documentclass[twocolumn,floatfix,showpacs,prd,aps,eqsecnum,tightenlines,superscriptaddress]{revtex4}
\usepackage{graphicx}
\usepackage{dcolumn}
\usepackage{bm}

\begin{document}


\title{Radiation content of Conformally flat initial data}

\author{C. O. Lousto} \affiliation{Department of Physics and Astronomy,
and Center for Gravitational Wave Astronomy,
The University of Texas at Brownsville, Brownsville, Texas 78520}
\author{Richard H. Price} \affiliation{Department of Physics, University of Utah,
Salt Lake City, UT 84112}

\date{\today}

\begin{abstract}
We study the radiation of energy and linear momentum emitted to
infinity by the headon collision of binary black holes, starting from
rest at a finite initial separation, in the extreme mass ratio limit.
For these configurations we identify the radiation produced by the
initially conformally flat choice of the three geometry. This
identification suggests that the radiated energy and momentum of
headon collisions will not be dominated by the details of the initial
data for evolution of holes from initial proper separations
$L_0\geq7M$.  For non-headon orbits, where the amount of radiation is
orders of magnitude larger, the conformally flat initial data may
provide a relative even better approximation.
\end{abstract}

 \maketitle

\section{Introduction}\label{Sec:Intro}

As we get closer to the first direct detection of gravitational waves,
there is a renewed interest in the construction of astrophysically
relevant initial data for binary black holes, {\it a priori} thought
to be the most violent source of gravitational radiation in the
Universe.  The traditional choice for initial data had been the
classical work of Bowen and York~\cite{Bowen80}.  More recently,
attention has been directed toward `thin-sandwich'
\cite{Gourgoulhon02,Hannam:2003tv} and Kerr-Schild {\em
ans\"atze}~\cite{Marronetti00,Bishop:1998my} initial data, and comparisons
with the Bowen-York choice have been carried out on  the initial
slices~\cite{Pfeiffer:2002xz,Tichy03a,Laguna:2003sr}. Post-Newtonian
limit inspired data are also beginning to appear in the
literature~\cite{Tichy02,Blanchet:2003kz}.

The most important question about the choice of initial data is the
effect of that choice on the gravitational waveform at
infinity. Answering that question is very difficult if the binary
black holes have comparable
mass~\cite{Baker00b,Baker:2001sf,Baker:2001nu,Baker:2002qf,Baker:2003ds},
but is relatively straightforward in the particle limit. For this
reason, we have previously studied the headon collision of binary
black holes in the extreme mass ratio
regime\cite{Lousto97a,Lousto97b,Lousto98a}. This allowed us to compare
`on slice' initial data and truly evolved
data. In~\cite{Lousto97b,Lousto98a} we focused particular attention on
the importance of the choice of the extrinsic curvature for the
initial data, and we considered only the usual choice for the initial
three metric: conformal flatness.  In this report we turn to the
question omitted from ~\cite{Lousto97b,Lousto98a}, and we look briefly at
the importance of conformal flatness. We
consider time symmetric initial data so that the natural choice for
the extrinsic curvature is for it to vanish.
We use the Brill-Lindquist type of conformally flat initial data.
(We have shown in Ref.~\cite{Lousto97b}  that the total radiated
energy  is not very sensitive to this choice.) We evolve the time
symmetric initial data, compute the radiated waveforms, and try to
identify what part of the radiation can be ascribed to the conformally
flat choice of the initial three metric.  We do this differently for
large and for small initial separation of the holes. In the case of
large initial separation, we can identify an early feature of the
waveform that is clearly produced by the initial data. For the case of
small initial separation, we introduce a more speculative measure of
the radiation ascribed to the conformally flat initial data; we
consider it to be the excess radiation above the minimum of a
one-parameter family of initial data choices.

\section{Results}\label{Sec:results}

We consider a particle of mass $m_0$, to be a first-order
perturbation on the background of a Schwarzschild hole of mass $M$.
The particle is initially at rest at a proper distance $L_0$
from the horizon of the Schwarzschild hole.  For the radial infall of
the particle, the odd parity perturbations of the spacetime
vanish. For the even parity perturbations we use the
formulae and numerical techniques described in~\cite{Lousto97b} to
integrate the Moncrief-Zerilli equation
\begin{equation}\label{Zeq}
\left[\partial^2_{r^*}-\partial^2_t-V_\ell(r)\right]\psi_\ell
=S_\ell(r_p(t),r)\,,
\end{equation}
for the even parity wave function $\psi_\ell(t,r)$ for each $\ell$-pole
mode of the field.

\subsection{radiated energies}

\begin{figure}
\begin{center}
\includegraphics[width=3.2in]{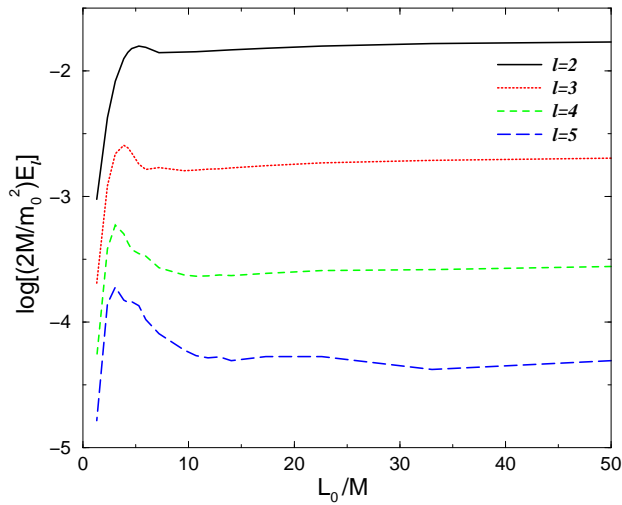}
\end{center}
\caption{The contribution to the radiated energy at infinity of the
first four multipoles, as a function of the initial location of 
the particle falling from rest.}
\label{fig:logEl}
\end{figure}

The radiated even-parity energy at infinity is computed from
\begin{equation}
\frac{dE}{dt}=\lim_{r\to\infty}\frac{1}{64\pi}\sum_{\ell=2}^\infty
\frac{(\ell+2)!}{(\ell-2)!}\left(\partial_t\psi_\ell(r,t)\right)^2\,.
\label{Edot}
\end{equation}
The computed energy $E_\ell$ for $\ell$=2\ldots5, reported in
Table~\ref{table:E} and Fig.~\ref{fig:logEl}, show the expected
exponential decrease of energy with increasing
$\ell$\cite{Davis71,Lousto97a}.
The plot also displays an
unexpected feature: For small separations there is a rise in the
radiated energy. 
The
relative importance of this feature grows with increasing $\ell$, and
for  $\ell\geq3$ the feature is so strong that
radiation from small $L_0/M$  is greater than that for infall
from infinity.

\begin{figure}
\begin{center}
\includegraphics[width=3.2in]{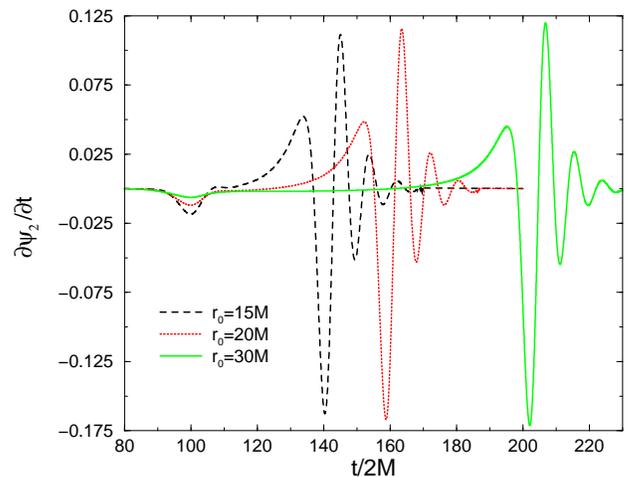}
\end{center}
\caption{Waveforms of $\partial\psi_2/\partial t$ at $r^*/2M=100$.
Note the small feature located around $t/2M=100$, associated with the
initial data radiation content.  }
\label{fig:dwvfr}
\end{figure}

\begin{figure}
\begin{center}
\includegraphics[width=3.2in]{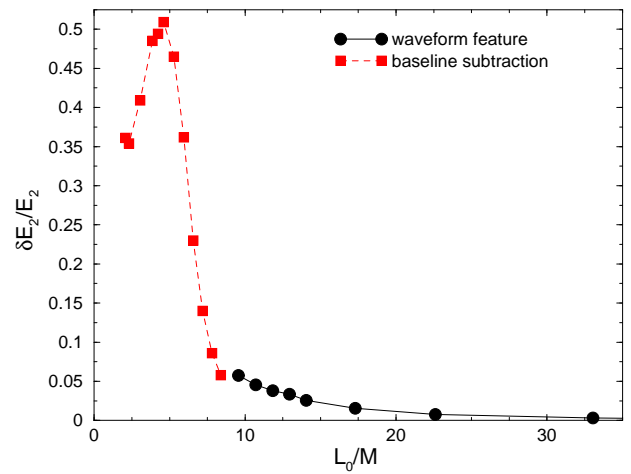}
\end{center}
\caption{The estimated fraction of radiated $\ell=2$ energy that 
is due to the choice of conformally flat initial data. The circles,
for $L_0/M>9.56$ represent fractional energy in the early feature in 
the waveform. The squares, for $L_0/M\leq8$, are based on the 
energy in excess of a baseline model of initial data that is 
not conformally flat.}
\label{fig:dE2}
\end{figure}

This increase in $E_\ell$ for small $L_0/M$ is certainly related to
the choice of initial data.  As a first step in understanding this
relationship, in Fig.~\ref{fig:dwvfr} we display
$\partial\psi_2/\partial t$, the quantity that dominates the energy in
Table~\ref{table:E} and Fig.~\ref{fig:logEl}.  The waveform of
$\partial\psi_2/\partial t$, for an observer at $r^*/2M=100$,
shows the characteristic excitation and ringdown of quasinormal
oscillations.  But at early times, $t/2M\sim100$, a small feature
appears representing the evolution of the initial disturbance of the
Schwarzschild background around the particle position.  We have
isolated this small feature and have computed its contribution $\delta
E_2$ to the quadrupole radiated energy $E_2$.  The values of $\delta
E_2$ are listed in Table~\ref{table:E}, and the circles in
Fig.~\ref{fig:dE2} show the fractional energy $\delta E_2/E_2$ as a
function of the initial separation.  The circles in Fig.~\ref{fig:dE2}
show a marked increase in the fractional energy as $L_0/M$ decreases,
but this method of ascribing energy to the initial data can only be
extended down to $L_0/M\sim10$. For $L_0/M<10$ the small early feature
in the waveform cannot be cleanly distinguished from the initial
excitation of ringing. (The error in $\delta E_2$ reaches several
percent for the values tabulated in Table~\ref{table:E}.)

One can make a speculative estimate of the small-$L_0$ initial-value
energy by extrapolation.  A best fit of the form $\delta
E_2/E_2=a(L_0/M)^{b}$ requires $a=7.26$ and $b=-2.135$.  This fitted
curve reaches $\delta E_2/E_2\approx0.5$ for $L_0/M\approx3.5$.

There is a completely different way to estimate the energy associated
with the conformally flat initial data.  To do this, we follow the
approach of Martel and Poisson\cite{Martel:2001yf}.  In the
Regge-Wheeler\cite{Regge57} gauge for the multipole perturbations of
Schwarzschild spacetime, the only nonvanishing even parity
perturbations on the $t=$\,constant initial hypersurface are $H_2^{\ell m}$
and $K^{\ell m}$. The  initial value equations 
contain a free functional degree of freedom in $H_2^{\ell m}$
and $K^{\ell m}$. The usual choice, $H_2^{\ell m}=K^{\ell m}$, corresponds
to a conformally flat perturbed three metric. Here, instead, we
consider the one parameter family
\begin{equation}\label{alphamodel}
H_2^\ell=\alpha K^\ell\,.
\end{equation}
In Ref.\cite{Martel:2001yf}, the energy radiated has been studied as a
function of $\alpha$, and the value of $\alpha$ has been found for
which the radiated energy is minimum. This turns out {\em not} to be
the conformally flat choice $\alpha=1$. (See Fig.~12 in
Ref.~\cite{Martel:2001yf}.)
For any value of 
$L_0/M$, 
 we take the $\alpha$-model of Eq.~(\ref{alphamodel}) with 
the minimum radiated energy as the ``baseline'' model.
We subtract the radiated energy for this baseline model
from the radiated energy for a
conformally flat model
and consider the excess energy to be an artifact of 
the conformally flat initial data, radiation energy that is, in 
a sense, contained within the initial data.
This excess energy, for $\ell=2$ and $L_0/M\leq8$, is included in
Fig.~\ref{fig:dE2}.  It may be interesting that 
the maximum of this excess energy is located
near a proper separation $L_0/M\approx4.5$, in rough agreement with
the previous extrapolation.

\begin{table}
\begin{widetext}
\caption{Energy radiated in units of ${m_0^2}/{2M}$}
\begin{ruledtabular}
\begin{tabular}{lllllll}\label{table:E}
$L_0/M$ & $E_{2} $ & $E_{3}$ & $E_{4}$  & $E_{5}$ & $E_{Total} $& $\delta E_{2}/E_{2}$\\ 
 \hline
104.3	& 0.0179&0.00214&0.000318&9.69E-05 	&0.0205 &.4922e-4	\\
63.8	& 0.0174&0.00207&0.000289&5.61E-05	&0.0198	&.3695e-3	\\
33.0	& 0.0164&0.00194&0.000262&4.20E-05	&0.0187	&.3062e-2	\\
22.6	& 0.0157&0.00185&0.000257&5.31E-05	&0.0179	&.7707e-2	\\
17.3	& 0.0151&0.00176&0.000245&5.30E-05	&0.0172	&.1536e-1	\\
14.0	& 0.0147&0.00169&0.000234&4.92E-05	&0.0167	&.2557e-1	\\
12.9	& 0.0145&0.00166&0.000237&5.28E-05	&0.0165	&.3344e-1	\\
11.8	& 0.0143&0.00165&0.000234&5.20E-05	&0.0162	&.3811e-1	\\
10.7	& 0.0142&0.00162&0.000232&5.39E-05	&0.0161	&.4549e-1	\\
9.56	& 0.0141&0.00160&0.000237&5.99E-05	&0.0160	&.5730e-1	\\
7.19	& 0.0140&0.00170&0.000271&8.11E-05	&0.0160	&0.140	\\
5.94	& 0.0154&0.00164&0.000336&0.000105	&0.0174	&0.362	\\
5.28	& 0.0157&0.00181&0.000350&0.000135	&0.0180	&0.465	\\
4.59	& 0.0150&0.00223&0.000378&0.000145	&0.0177	&0.509	\\
4.23	& 0.0140&0.00244&0.000429&0.000144	&0.0170	&0.494	\\
3.86	& 0.0125&0.00256&0.000505&0.000150	&0.0157	&0.485	\\
3.05	& 0.00824&0.00219&0.000592&0.000189	&0.0112	&0.409	\\
2.30	& 0.00424&0.00122&0.000383&0.000140	&0.00595&0.354	\\
1.29	& 0.00095&0.00020&5.24E-05&1.64E-05	&0.00121&-----	
\end{tabular}
\end{ruledtabular}
\end{widetext} 
\end{table}

\subsection{radiated linear momentum}

An astrophysically interesting quantity to compute is the radiated
linear momentum and the corresponding recoil velocity of the final
black hole. If the recoil velocity can be comparable to the escape
velocity of a typical galaxy ($100-1000$\,km/s), or a cluster
($10-20$\,km/s),  recoil-driven escape can affect the growth rate
of massive black holes in the core of such star systems.  The
computation of this quantity for a particle released from rest at
infinity in a Kerr background was performed in
Refs.~\cite{Nakamura:1983hk,Kojima:1984cc}. For collisions of
comparable mass holes the radiated linear momentum in a headon
collision was computed in Ref.~\cite{Andrade:1997pc} in the close
limit, and for several larger initial separations in full 2D nonlinear
general relativity in Ref.~\cite{Anninos98a}. We present here explicit
results of the extreme mass ratio limit in order to provide
benchmarking data for future full nonlinear numerical work.

The radiated linear momentum along the axis of the collision is computed
as the correlation of two successive multipole
contributions~\cite{Moncrief80a}
\begin{equation}
\frac{dP_z}{dt}=\lim_{r\to\infty}\frac{1}{32\pi}\sum_{\ell=2}^\infty
\frac{(\ell+3)!}{(\ell-2)!}
\frac{\partial_t\psi_\ell(r,t)\,\partial_t\psi_{\ell+1}(r,t)}{\sqrt{(2\ell+1)(2\ell+3)}}\,.
\label{Pdot}
\end{equation}

\begin{figure}
\begin{center}
\includegraphics[width=3.2in]{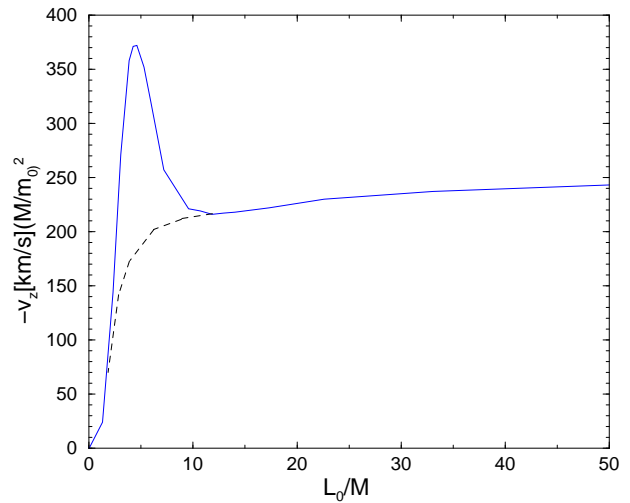}
\end{center}
\caption{Recoil velocity of the system for different initial proper
separations of the holes. The dashed line indicates an estimate of
recoil velocity for a baseline model of initial data that is not
conformally flat.}
\label{fig:P}
\end{figure}

Figure~\ref{fig:P} displays the results of computing the recoil
velocity of the binary system by summing over $\ell=2\cdots5$. A
notable feature here is that radiated momentum is a maximum for a
particle that starts at rest from $L_0/M\sim4.5$, not from infinity.
For this momentum maximum, the recoil velocity reaches almost
400$(m_0/M)^2$\,km/s compared to 250 $(m_0/M)^2$\,km/s for a particle
from infinity (See Table~\ref{table:P}).  As we did with the radiated
energy, we can make a speculative estimate of the fraction of the
recoil velocity that may be ascribed to the conformally flat initial
data. We take $\delta v$ to be the excess recoil velocity above that
of the baseline (energy minimum) model.  At the peak recoil velocity
we find that $\delta v$ is 1.0 times the recoil velocity for the
baseline model. Thus the part of the recoil that we view as an
artifact of the initial data is as big as the baseline value. This effect is
more pronounced than that for energy because the radiation of momentum
involves  $\ell=3$ as much as it does $\ell=2$; as shown in
Fig.~\ref{fig:logEl}, the sensitivity of $\partial\psi_\ell/\partial
t$ to initial data increases with increasing $\ell$.

\begin{table}
\begin{widetext}
\caption{Linear momentum radiated}
\begin{ruledtabular}
\begin{tabular}{llc}\label{table:P}
$r_0/M$	&	$L_0/M$ & $(M/m_0)^2v$ km/s \\ 
\hline
100	&104.3	& 249	\\
60	&63.8	& 248	\\
30	&33.0	& 237	\\
20	&22.6	& 230	\\
15	&17.3	& 222	\\
12	&14.0	& 218	\\
11	&12.9	& 217	\\
10	&11.8	& 216	\\
9	&10.7	& 219	\\
8	&9.56	& 221	\\
6	&7.19	& 257	\\
5	&5.94	& 320	\\
4.5	&5.28	& 352	\\
4	&4.59	& 372	\\
3.75	&4.23	& 371	\\
3.5	&3.86	& 358	\\
3	&3.05	& 271	\\
2.6	&2.30	& 144	\\
2.2	&1.29	& 24	\\
\end{tabular}
\end{ruledtabular}
\end{widetext} 
\end{table}


\section{Discussion}

To assess the importance of the choice of the initial three geometry
on the computation of radiation from black hole mergers, we have used
the very simplest model: the extreme mass limit, a nonrotating hole,
and radial infall from rest.  Our results suggest that for proper
distance $L_0$ larger than around $7M$ the energy and momentum
radiated are probably not significantly contaminated by the initial
choice.  In the astrophysically more interesting case of inspiralling
binary black holes in quasicircular orbits, we expect the radiation
will be affected much more by the choice of the extrinsic curvature,
than the choice of the three metric. In that case, the conformally
flat choice of the three-geometry seems adequate for most of the
applications. A study of the initial data choices for quasicircular
orbits in the small mass ratio limit is currently underway by the
authors.


\acknowledgments 
We gratefully acknowledge the support of the
NASA Center for Gravitational Wave Astronomy at University of Texas at
Brownsville (NAG5-13396). RHP thanks the National Science Foundation
for support under grant PHY0244605. And C.O.L for financial support
from grant NSF-PHY-0140326

\bibliographystyle{apsrev}
\bibliography{bibtex/references}

\begin{thebibliography}{26}
\expandafter\ifx\csname natexlab\endcsname\relax\def\natexlab#1{#1}\fi
\expandafter\ifx\csname bibnamefont\endcsname\relax
  \def\bibnamefont#1{#1}\fi
\expandafter\ifx\csname bibfnamefont\endcsname\relax
  \def\bibfnamefont#1{#1}\fi
\expandafter\ifx\csname citenamefont\endcsname\relax
  \def\citenamefont#1{#1}\fi
\expandafter\ifx\csname url\endcsname\relax
  \def\url#1{\texttt{#1}}\fi
\expandafter\ifx\csname urlprefix\endcsname\relax\def\urlprefix{URL }\fi
\providecommand{\bibinfo}[2]{#2}
\providecommand{\eprint}[2][]{\url{#2}}

\bibitem[{\citenamefont{Bowen and York}(1980)}]{Bowen80}
\bibinfo{author}{\bibfnamefont{J.}~\bibnamefont{Bowen}} \bibnamefont{and}
  \bibinfo{author}{\bibfnamefont{J.~W.} \bibnamefont{York}},
  \bibinfo{journal}{Phys. Rev. D} \textbf{\bibinfo{volume}{21}},
  \bibinfo{pages}{2047} (\bibinfo{year}{1980}).

\bibitem[{\citenamefont{Gourgoulhon et~al.}(2002)\citenamefont{Gourgoulhon,
  Grandclement, and Bonazzola}}]{Gourgoulhon02}
\bibinfo{author}{\bibfnamefont{E.}~\bibnamefont{Gourgoulhon}},
  \bibinfo{author}{\bibfnamefont{P.}~\bibnamefont{Grandclement}},
  \bibnamefont{and}
  \bibinfo{author}{\bibfnamefont{S.}~\bibnamefont{Bonazzola}},
  \bibinfo{journal}{Phys. Rev. D} \textbf{\bibinfo{volume}{65}},
  \bibinfo{pages}{044020} (\bibinfo{year}{2002}), \eprint{gr-qc/0106015}.

\bibitem[{\citenamefont{Hannam et~al.}(2003)\citenamefont{Hannam, Evans, Cook,
  and Baumgarte}}]{Hannam:2003tv}
\bibinfo{author}{\bibfnamefont{M.~D.} \bibnamefont{Hannam}},
  \bibinfo{author}{\bibfnamefont{C.~R.} \bibnamefont{Evans}},
  \bibinfo{author}{\bibfnamefont{G.~B.} \bibnamefont{Cook}}, \bibnamefont{and}
  \bibinfo{author}{\bibfnamefont{T.~W.} \bibnamefont{Baumgarte}}
  (\bibinfo{year}{2003}), \eprint{gr-qc/0306028}.

\bibitem[{\citenamefont{Marronetti and Matzner}(2000)}]{Marronetti00}
\bibinfo{author}{\bibfnamefont{P.}~\bibnamefont{Marronetti}} \bibnamefont{and}
  \bibinfo{author}{\bibfnamefont{R.~A.} \bibnamefont{Matzner}},
  \bibinfo{journal}{Phys. Rev. Lett.} \textbf{\bibinfo{volume}{85}},
  \bibinfo{pages}{5500} (\bibinfo{year}{2000}), \bibinfo{note}{gr-qc/0009044}.

\bibitem[{\citenamefont{Bishop et~al.}(1998)\citenamefont{Bishop, Isaacson,
  Maharaj, and Winicour}}]{Bishop:1998my}
\bibinfo{author}{\bibfnamefont{N.~T.} \bibnamefont{Bishop}},
  \bibinfo{author}{\bibfnamefont{R.}~\bibnamefont{Isaacson}},
  \bibinfo{author}{\bibfnamefont{M.}~\bibnamefont{Maharaj}}, \bibnamefont{and}
  \bibinfo{author}{\bibfnamefont{J.}~\bibnamefont{Winicour}},
  \bibinfo{journal}{Phys. Rev.} \textbf{\bibinfo{volume}{D57}},
  \bibinfo{pages}{6113} (\bibinfo{year}{1998}), \eprint{gr-qc/9711076}.

\bibitem[{\citenamefont{Pfeiffer et~al.}(2002)\citenamefont{Pfeiffer, Cook, and
  Teukolsky}}]{Pfeiffer:2002xz}
\bibinfo{author}{\bibfnamefont{H.~P.} \bibnamefont{Pfeiffer}},
  \bibinfo{author}{\bibfnamefont{G.~B.} \bibnamefont{Cook}}, \bibnamefont{and}
  \bibinfo{author}{\bibfnamefont{S.~A.} \bibnamefont{Teukolsky}},
  \bibinfo{journal}{Phys. Rev.} \textbf{\bibinfo{volume}{D66}},
  \bibinfo{pages}{024047} (\bibinfo{year}{2002}), \eprint{gr-qc/0203085}.

\bibitem[{\citenamefont{Tichy et~al.}(2003{\natexlab{a}})\citenamefont{Tichy,
  Br\"ugmann, and Laguna}}]{Tichy03a}
\bibinfo{author}{\bibfnamefont{W.}~\bibnamefont{Tichy}},
  \bibinfo{author}{\bibfnamefont{B.}~\bibnamefont{Br\"ugmann}},
  \bibnamefont{and} \bibinfo{author}{\bibfnamefont{P.}~\bibnamefont{Laguna}},
  \bibinfo{journal}{Phys. Rev. D} \textbf{\bibinfo{volume}{68}},
  \bibinfo{pages}{064008} (\bibinfo{year}{2003}{\natexlab{a}}),
  \bibinfo{note}{gr-qc/0306020}.

\bibitem[{\citenamefont{Laguna}(2003)}]{Laguna:2003sr}
\bibinfo{author}{\bibfnamefont{P.}~\bibnamefont{Laguna}}
  (\bibinfo{year}{2003}), \eprint{gr-qc/0310073}.

\bibitem[{\citenamefont{Tichy et~al.}(2003{\natexlab{b}})\citenamefont{Tichy,
  Br\"ugmann, Campanelli, and Diener}}]{Tichy02}
\bibinfo{author}{\bibfnamefont{W.}~\bibnamefont{Tichy}},
  \bibinfo{author}{\bibfnamefont{B.}~\bibnamefont{Br\"ugmann}},
  \bibinfo{author}{\bibfnamefont{M.}~\bibnamefont{Campanelli}},
  \bibnamefont{and} \bibinfo{author}{\bibfnamefont{P.}~\bibnamefont{Diener}},
  \bibinfo{journal}{Phys. Rev. D} \textbf{\bibinfo{volume}{67}},
  \bibinfo{pages}{064008} (\bibinfo{year}{2003}{\natexlab{b}}),
  \bibinfo{note}{gr-qc/0207011}.

\bibitem[{\citenamefont{Blanchet}(2003)}]{Blanchet:2003kz}
\bibinfo{author}{\bibfnamefont{L.}~\bibnamefont{Blanchet}}
  (\bibinfo{year}{2003}), \eprint{gr-qc/0304080}.

\bibitem[{\citenamefont{Baker et~al.}(2000)\citenamefont{Baker, Br\"ugmann,
  Campanelli, and Lousto}}]{Baker00b}
\bibinfo{author}{\bibfnamefont{J.}~\bibnamefont{Baker}},
  \bibinfo{author}{\bibfnamefont{B.}~\bibnamefont{Br\"ugmann}},
  \bibinfo{author}{\bibfnamefont{M.}~\bibnamefont{Campanelli}},
  \bibnamefont{and} \bibinfo{author}{\bibfnamefont{C.~O.}
  \bibnamefont{Lousto}}, \bibinfo{journal}{Class. Quantum Grav.}
  \textbf{\bibinfo{volume}{17}}, \bibinfo{pages}{L149} (\bibinfo{year}{2000}).

\bibitem[{\citenamefont{Baker et~al.}(2002{\natexlab{a}})\citenamefont{Baker,
  Campanelli, and Lousto}}]{Baker:2001sf}
\bibinfo{author}{\bibfnamefont{J.}~\bibnamefont{Baker}},
  \bibinfo{author}{\bibfnamefont{M.}~\bibnamefont{Campanelli}},
  \bibnamefont{and} \bibinfo{author}{\bibfnamefont{C.~O.}
  \bibnamefont{Lousto}}, \bibinfo{journal}{Phys. Rev.}
  \textbf{\bibinfo{volume}{D65}}, \bibinfo{pages}{044001}
  (\bibinfo{year}{2002}{\natexlab{a}}),
  \eprint[http://arXiv.org/abs]{gr-qc/0104063}.

\bibitem[{\citenamefont{Baker et~al.}(2001)\citenamefont{Baker, Br{\"u}gmann,
  Campanelli, Lousto, and Takahashi}}]{Baker:2001nu}
\bibinfo{author}{\bibfnamefont{J.}~\bibnamefont{Baker}},
  \bibinfo{author}{\bibfnamefont{B.}~\bibnamefont{Br{\"u}gmann}},
  \bibinfo{author}{\bibfnamefont{M.}~\bibnamefont{Campanelli}},
  \bibinfo{author}{\bibfnamefont{C.~O.} \bibnamefont{Lousto}},
  \bibnamefont{and}
  \bibinfo{author}{\bibfnamefont{R.}~\bibnamefont{Takahashi}},
  \bibinfo{journal}{Phys. Rev. Lett.} \textbf{\bibinfo{volume}{87}},
  \bibinfo{pages}{121103} (\bibinfo{year}{2001}),
  \eprint[http://arXiv.org/abs]{gr-qc/0102037}.

\bibitem[{\citenamefont{Baker et~al.}(2002{\natexlab{b}})\citenamefont{Baker,
  Campanelli, Lousto, and Takahashi}}]{Baker:2002qf}
\bibinfo{author}{\bibfnamefont{J.}~\bibnamefont{Baker}},
  \bibinfo{author}{\bibfnamefont{M.}~\bibnamefont{Campanelli}},
  \bibinfo{author}{\bibfnamefont{C.~O.} \bibnamefont{Lousto}},
  \bibnamefont{and}
  \bibinfo{author}{\bibfnamefont{R.}~\bibnamefont{Takahashi}},
  \bibinfo{journal}{Phys. Rev.} \textbf{\bibinfo{volume}{D65}},
  \bibinfo{pages}{124012} (\bibinfo{year}{2002}{\natexlab{b}}),
  \eprint[http://arXiv.org/abs]{astro-ph/0202469}.

\bibitem[{\citenamefont{Baker et~al.}(2003)\citenamefont{Baker, Campanelli,
  Lousto, and Takahashi}}]{Baker:2003ds}
\bibinfo{author}{\bibfnamefont{J.}~\bibnamefont{Baker}},
  \bibinfo{author}{\bibfnamefont{M.}~\bibnamefont{Campanelli}},
  \bibinfo{author}{\bibfnamefont{C.~O.} \bibnamefont{Lousto}},
  \bibnamefont{and} \bibinfo{author}{\bibfnamefont{R.}~\bibnamefont{Takahashi}}
  (\bibinfo{year}{2003}), \eprint[http://arXiv.org/abs]{astro-ph/0305287}.

\bibitem[{\citenamefont{Lousto and Price}(1997{\natexlab{a}})}]{Lousto97a}
\bibinfo{author}{\bibfnamefont{C.~O.} \bibnamefont{Lousto}} \bibnamefont{and}
  \bibinfo{author}{\bibfnamefont{R.~H.} \bibnamefont{Price}},
  \bibinfo{journal}{Phys. Rev.} \textbf{\bibinfo{volume}{D55}},
  \bibinfo{pages}{2124} (\bibinfo{year}{1997}{\natexlab{a}}),
  \eprint{gr-qc/9609012}.

\bibitem[{\citenamefont{Lousto and Price}(1997{\natexlab{b}})}]{Lousto97b}
\bibinfo{author}{\bibfnamefont{C.~O.} \bibnamefont{Lousto}} \bibnamefont{and}
  \bibinfo{author}{\bibfnamefont{R.~H.} \bibnamefont{Price}},
  \bibinfo{journal}{Phys. Rev.} \textbf{\bibinfo{volume}{D56}},
  \bibinfo{pages}{6439} (\bibinfo{year}{1997}{\natexlab{b}}),
  \eprint{gr-qc/9705071}.

\bibitem[{\citenamefont{Lousto and Price}(1998)}]{Lousto98a}
\bibinfo{author}{\bibfnamefont{C.~O.} \bibnamefont{Lousto}} \bibnamefont{and}
  \bibinfo{author}{\bibfnamefont{R.~H.} \bibnamefont{Price}},
  \bibinfo{journal}{Phys. Rev.} \textbf{\bibinfo{volume}{D57}},
  \bibinfo{pages}{1073} (\bibinfo{year}{1998}), \eprint{gr-qc/9708022}.

\bibitem[{\citenamefont{Davis et~al.}(1971)\citenamefont{Davis, Ruffini, Press,
  and Price}}]{Davis71}
\bibinfo{author}{\bibfnamefont{M.}~\bibnamefont{Davis}},
  \bibinfo{author}{\bibfnamefont{R.}~\bibnamefont{Ruffini}},
  \bibinfo{author}{\bibfnamefont{H.}~\bibnamefont{Press}}, \bibnamefont{and}
  \bibinfo{author}{\bibfnamefont{R.~H.} \bibnamefont{Price}},
  \bibinfo{journal}{Phys. Rev. Lett.} \textbf{\bibinfo{volume}{27}},
  \bibinfo{pages}{1466} (\bibinfo{year}{1971}).

\bibitem[{\citenamefont{Martel and Poisson}(2002)}]{Martel:2001yf}
\bibinfo{author}{\bibfnamefont{K.}~\bibnamefont{Martel}} \bibnamefont{and}
  \bibinfo{author}{\bibfnamefont{E.}~\bibnamefont{Poisson}},
  \bibinfo{journal}{Phys. Rev. D} \textbf{\bibinfo{volume}{66}},
  \bibinfo{pages}{084001} (\bibinfo{year}{2002}),
  \bibinfo{note}{gr-qc/0107104}.

\bibitem[{\citenamefont{Regge and Wheeler}(1957)}]{Regge57}
\bibinfo{author}{\bibfnamefont{T.}~\bibnamefont{Regge}} \bibnamefont{and}
  \bibinfo{author}{\bibfnamefont{J.}~\bibnamefont{Wheeler}},
  \bibinfo{journal}{Phys. Rev.} \textbf{\bibinfo{volume}{108}},
  \bibinfo{pages}{1063} (\bibinfo{year}{1957}).

\bibitem[{\citenamefont{Nakamura and Haugan}(1983)}]{Nakamura:1983hk}
\bibinfo{author}{\bibfnamefont{T.}~\bibnamefont{Nakamura}} \bibnamefont{and}
  \bibinfo{author}{\bibfnamefont{M.~P.} \bibnamefont{Haugan}},
  \bibinfo{journal}{Astrophys. J.} \textbf{\bibinfo{volume}{269}},
  \bibinfo{pages}{292} (\bibinfo{year}{1983}).

\bibitem[{\citenamefont{Kojima and Nakamura}(1984)}]{Kojima:1984cc}
\bibinfo{author}{\bibfnamefont{Y.}~\bibnamefont{Kojima}} \bibnamefont{and}
  \bibinfo{author}{\bibfnamefont{T.}~\bibnamefont{Nakamura}},
  \bibinfo{journal}{Prog. Theor. Phys.} \textbf{\bibinfo{volume}{72}},
  \bibinfo{pages}{494} (\bibinfo{year}{1984}).

\bibitem[{\citenamefont{Andrade and Price}(1997)}]{Andrade:1997pc}
\bibinfo{author}{\bibfnamefont{Z.}~\bibnamefont{Andrade}} \bibnamefont{and}
  \bibinfo{author}{\bibfnamefont{R.~H.} \bibnamefont{Price}},
  \bibinfo{journal}{Phys. Rev.} \textbf{\bibinfo{volume}{D56}},
  \bibinfo{pages}{6336} (\bibinfo{year}{1997}), \eprint{gr-qc/9611022}.

\bibitem[{\citenamefont{Anninos and Brandt}(1998)}]{Anninos98a}
\bibinfo{author}{\bibfnamefont{P.}~\bibnamefont{Anninos}} \bibnamefont{and}
  \bibinfo{author}{\bibfnamefont{S.}~\bibnamefont{Brandt}},
  \bibinfo{journal}{Phys. Rev. Lett.} \textbf{\bibinfo{volume}{81}},
  \bibinfo{pages}{508} (\bibinfo{year}{1998}), \eprint{gr-qc/9806031}.

\bibitem[{\citenamefont{Moncrief}(1980)}]{Moncrief80a}
\bibinfo{author}{\bibfnamefont{V.}~\bibnamefont{Moncrief}},
  \bibinfo{journal}{Astrophys.J.} \textbf{\bibinfo{volume}{238}},
  \bibinfo{pages}{333} (\bibinfo{year}{1980}).

\end{thebibliography}
\thebibliography{momento2}
\end{document}